\documentclass[12pt]{article}
\usepackage[utf8]{inputenc}
\usepackage[hscale=0.76,vscale=0.82]{geometry}
\usepackage[at]{easylist}
\usepackage{graphicx}
\usepackage{caption}
\usepackage{subcaption}
\usepackage{amsmath}
\usepackage{float}
\usepackage{hanging}
\restylefloat{table}
\counterwithin{equation}{section}
\title{Bootstrap Confidence Intervals Using the Likelihood Ratio Test in Changepoint Detection}
\author{Ryan Chen - under the supervision of \\ Professor Javier Cabrera at Rugters University}
\date{April 29, 2020}

\begin{document}
\maketitle
\newpage

\begin{abstract}
    This study aims to evaluate the performance of power in the likelihood ratio test for changepoint detection by bootstrap sampling, and proposes a hypothesis test based on bootstrapped confidence interval lengths. Assuming i.i.d normally distributed errors, and using the bootstrap method, the changepoint sampling distribution is estimated. Furthermore, this study describes a method to estimate a data set with no changepoint to form the null sampling distribution. With the null sampling distribution, and the distribution of the estimated changepoint, critical values and power calculations can be made, over the lengths of confidence intervals. 
\end{abstract}

\section{Introduction}
Changepoint detection has been heavily studied in statistics and engineering. There are many ways to determine the location of changepoints accurately and quickly, summarised by Aminikhanghahi and Cook (2016). One of the simpler methods of changepoint detection derived from foundations of hypothesis testing and linear models employs the likelihood ratio test. Typically the data is assumed to be normally distributed. Kim and Siegmund (1989) have discussed extensively on the theoretical form of the changepoint test statistic and its sampling distribution, under these assumptions. Yet a problem remains as Hušková and Kirsch (2011) pointed out, that often times, there are not enough time and resources to gather enough data points to achieve a desired power. Thus they sought a bootstrap approach.

The use of bootstrap to evaluate changepoints has been discussed by Kirsch (2008). The analysis by Hušková and Kirsch (2018)  also explored how bootstrap confidence bands change in length based on factors such as a change in the mean at the changepoint. Their study visits the change in mean over the unit scale, though this may be highly variable under data sets where there is greater variance. This analysis draws on their idea of bootstrapping and its inference, and analyzes the power of changepoint detection by viewing the effect size as a function of the variance.

First, a bootstrap method is proposed in evaluating whether a changepoint exists and uses bootstrap samples in an effort to identify the distributions related to the changepoint. Suppose $x_1\hdots x_n$ are independent random normal observations. To answer the question of whether a changepoint exists, the hypothesis of interest is set up as the following:

Let the linearly constant mean for $x_1\hdots x_k$ for $k<n$ be denoted by $\mu_1$ which takes the form of $\beta_{0,1} + t\beta_{1,1}$, and another linearly constant mean for $x_{k+1}...x_n$ be denoted by $\mu_2$ which takes the form of $\beta_{0,2} + t\beta_{1,2}$. The null hypothesis states that there is no changepoint, which can be expressed as $\mu_1=\mu_2=\mu$. The alternate hypothesis states that there is a changepoint at $k$ expressed as $\mu_1 \neq \mu_2$. That is, the question of whether or not a changepoint exists can be answered by testing $H_0: \mu_1 = \mu_2 (= \mu)$ against $H_\alpha: \mu_1\neq \mu_2$. Other variations of the hypothesis exist, such as testing if the changepoint occurs before the end of the data set versus if it occurs at the end of the data stream, as demonstrated in Hušková and Picek (2005).

\section{Likelihood Ratio Test}

The likelihood ratio test for changepoint detection, explored by Kim and Siegmund, is explained thoroughly by Dette and Gosmann (2019). In analyzing at-most-one-changepoint data, a term coined by Hušková and Kirsch (2018), there is the assumption that there could be at most two means in the data. For simplicity, this study uses normally distributed data, which is fit with ordinary least squares regression. 

The likelihood ratio test looks at the error distribution of two models, fit for two separate means, split at each observation $x_i$, sequentially for each $i=1\hdots n$. It then compares the likelihood to that of the model which is fit on the whole data set, which assumes one mean and no changepoint. The test statistic $T$ evaluated at each observation $i$ can be given by:
\begin{align}
    T^C = \frac{\prod_{i=1}^{k}f_1(x_i)\prod_{i=k+1}^{n}f_2(x_i)}{\prod_{i=1}^n f(x_1)}
\end{align}

Here $f_i(x_i)$ represents the likelihood function of the mean up to observation $x_i$. The value of $k$ that maximizes $T^C$ is considered the changepoint. Thus by using log likelihoods, the changepoint $c_0$ can be estimated by solving:
\begin{align}
    \hat{c}_0 = \max_k \{\sum_{i=1}^k \log(f_1(x_i)) + \sum_{i=k+1}^n \log(f_2(x_i)) - \sum_{i=1}^n \log(f(x_i)) \}
\end{align}
The solution can be solved easily through computational methods, by using the log likelihood in (2.2) since empirically, the log likelihood function for the normal distribution is only a function of the number of observations going into the likelihood, and the maximum likelihood estimate of the variance $\hat{\sigma}^2$. Thus, in the bivariate case, using simple linear regression, the log likelihood method can identify changepoints where there are differences in means and trends. 

\section{Inference about the changepoint}

While there is numerous literature on the theoretical sampling distribution of the changepoint, another way to extract confidence intervals for changepoint estimates is by using the bootstrap method. The bootstrap method is shown to converge in probability to the sample distribution of interest in Singh (1981), so for a large enough bootstrap sample size, the changepoint sampling distribution can be estimated. From the sampling distribution, confidence bands of any level $\alpha$ can be extracted by: 
\begin{align}
    (\hat{\theta}^* - \hat{\theta}^*_{1-\alpha/2}, \hat{\theta}^* - \hat{\theta}^*_{\alpha/2})
\end{align}
Thus, the sampling distribution of the changepoint can be computed through several bootstrap samples of the data. In order to preserve the structure of the original data set, the bootstrap samples must be sampled in a way that preserves the sequential order of the data, that is the $(x_i, y_i)$ pairs must always remain together for all $i$. So, the bootstrap sample in this study was formed by sampling values on one dimension and their corresponding values were matched accordingly. The likelihood ratio test can then be performed on each bootstrap sample to form the bootstrap estimate of sampling distribution the alternative hypothesis for changepoint. The expression (3.1) provides a means to evaluate the confidence bands from the bootstrap sampling distribution. Large confidence bands suggest less certainty in the location of the changepoint, while smaller confidence bands suggest more certainty. Thus, in a data set where the changepoint is very clear, a smaller bootstrap confidence interval is expected, and likewise in a data set where the changepoint is more obfuscated, the bootstrap confidence interval is expected to be wider.

This leads to another set of hypotheses that can be proposed for detecting whether or not a changepoint exists, which is based on the length of the confidence bands. In a data set with no changepoint at all, the confidence bands are expected to be wide. That is, the hypotheses test can be written as:
\begin{align}
    H_0: \frac{\lambda_1}{\lambda_0} = 1 \text{ against } H_\alpha: \frac{\lambda_1}{\lambda_0} \leq q
\end{align}
Here, $\lambda_1$ and $\lambda_0$ denote the length of the confidence interval for a data set with one changepoint, and the length of a confidence interval for a data set without a changepoint, respectively. The constant $q$ is a critical value that is less than 1, in which the ratio $\frac{\lambda_1}{\lambda_0}$ would need to fall under, in order to reject $H_0$, that there is no changepoint. If the confidence bands produced under the null and alternative sampling distributions are similar, then it is no evidence suggesting that there is a changepoint, and $H_0$ would fail to be rejected.

With a given data set $\textbf{x}=\{x_1 \hdots x_n\}$ where the changepoint $c_0$ and the magnitude of the change $\delta$ are initially unknown, the sampling distributions for the changepoint $c_0$ and $\delta$ must be estimated through the \textbf{x} along with $\lambda_0$ and $\lambda_1$. First, $c_0$ can be estimated  through equation (2.2), and the means on both sides of the changepoint $\mu_1$ and $\mu_2$ can be estimated by MLE using the likelihood functions $f_1(x_i)$ and $f_2(x_1)$. That is, under the normally distributed error assumption and linear means,
\begin{align}
    \hat{\mu}_i = \max_{\beta_{0,i}, \beta_{1,i}} (2\pi\sigma_i^2)^{\frac{-n_i}{2}}e^{\frac{\sum_{j=1}^{n_i}(y_j - \beta_{0,i} - \beta{1,i}x_j)^2}{2\sigma_i^2}} \text{ for i = 1,2}
\end{align}

Given $\hat{\mu}_1$ and $\hat{\mu}_2$, the estimate of $\delta$ can be given by $\hat{\delta}=\hat{\mu}_2 - \hat{\mu}_1$. In order to estimate the sampling distribution of $\lambda_0$ by bootstrap, there must first be an estimate of a data set with no changepoint denoted here as $\textbf{x*}$. This data set $\textbf{x*}$ can be imputed the following way, where $x_i$ is from the original data set $\textbf{x}$:
\begin{align}
    \textbf{x*} = \begin{cases}
    x_i & \text{ if } i \leq \hat{c}_0 \\
    x_i - \hat{\delta}  & \text{ if } i > \hat{c}_0
    \end{cases}
\end{align}

This new data set $\textbf{x*}$ is a data set with no changepoint, as it has been demeaned, and is able to provide a bootstrap sampling distribution for $\lambda_0$.

Several bootstrap samples for changepoint can be made using $\textbf{x*}$ and several bootstrap samples for the changepoint can be made with $\textbf{x}$. These bootstrap samples will form a distribution of $c_0$ under the null hypothesis and alternative hypothesis. From these distributions, confidence interval length $\lambda_0$ and $\lambda_1$ can be estimated by using expression (3.1) to find the confidence interval on the sampling distributions of $c_0$. Repeating this step several times will form the bootstrapped sampling distribution of $\lambda_0$ and $\lambda_1$.

From these distributions, it is easy to take the $\alpha$-level quantile, $t*$, of the $\lambda_0$ sampling distribution. Any confidence interval length generated from $\textbf{x}$ which falls below $t*$ is in the rejection region and is evidence suggesting a changepoint does exist. This provides a set of hypotheses, equivalent to (3.2):
\begin{align}
    H_0: \lambda_1 = \lambda_0 \text{ against } H_\alpha: \lambda_1 \leq q\lambda_0
\end{align}
That is, if the confidence interval length, under the assumption that there is a changepoint, is shorter than the confidence interval length under the the assumption that there is no changepoint, then there is evidence suggesting a changepoint exists, provided that the changepoint confidence interval length is short enough. 

The power can be calculated by taking the critical confidence interval length of $t*$ and seeing what proportions in the alternate distribution lie below $t*$. The assumption that the data set has a variance $\sigma^2$ presents an issue in analyzing the effect size on power. If the effect size is analyzed based on distance on multiples of $\sigma$, it would be easier to see how noisy data can affect the power of this method. Thus the figure below shows the way that power changes over effect size as a function of standard deviation. 

\begin{figure}[htp]
\centering
  \includegraphics[width=0.5\linewidth]{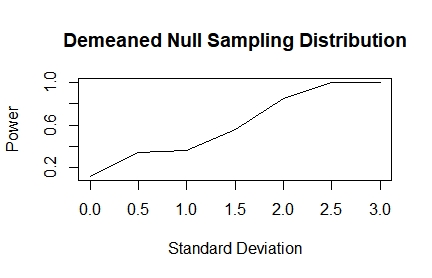}
  \caption{The power is very high for differences that are twice the standard deviation of the data set. $\alpha = 0.05$ with 1000 bootstrap samples}
  \label{fig:demean}
\end{figure}

\section{Discussion}

Another way to evaluate the null sampling distribution is by permuting one dimension of the data set, such that the pairs $(x_i,y_i)$ do not have to remain together. This shuffling removes any structure within the data set, if any, and thus should not contain any changepoint. Thus, it provides an alternative way to estimate the data set with no changepoint, that can then be applied to the same hypothesis test as in (3.5). 

However, it should be noted that these two methods would provide different results. By demeaning the data using $\hat{\delta}$, the residual structure still remains, from the original observed data set. By using this modification as an estimate for the data set with no changepoint, an assumption is made; that the points are merely shifted. This is analogous to overfitting, since the structure from the observed data is assumed to be the same structure in a data set with no changepoint. On the other hand, permuting the data would be analogous to underfitting an estimate of the no changepoint data set, as permutations completely shuffle away any structure in the data set. Thus, it is reasonable to expect the power achieved in tests that demean $\hat{\delta}$ to be considerably higher than the true power level ($1-\beta$). Similarly, the power achieved through permuting the data is expected to be considerably lower than the true power level. This is evident in the power curves in Figure 2. The power curve in Figure 2a is consistently higher than that in 2b. At $\delta=0$, the power from demeaning the data is 0.12, while the power from permuting the data is 0.02. This is in line with what is expected from the ``overfitted" and ``underfitted" null data set.

\begin{figure}[htp]
\centering
\begin{subfigure}{.45\textwidth}
  \centering
  \includegraphics[width=.85\linewidth]{demean.jpeg}
  \caption{Null data set by demeaning $\hat{\delta}$}
  \label{fig:sub1}
\end{subfigure}%
\begin{subfigure}{.45\textwidth}
  \centering
  \includegraphics[width=.85\linewidth]{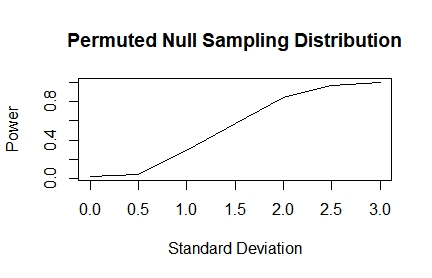}
  \caption{Null data set by permutation}
  \label{fig:sub2}
\end{subfigure}
\caption{Power Curves for Effect Size as Function of Standard Deviation at $\alpha=0.05$, 1000 bootstrap samples. At a 3 standard deviation difference, the power is very high.}
\label{fig:test}
\end{figure}
\section{Conclusion}

At the core of much of the literature on changepoint detection, much effort has been spent investigating the theoretical distributions and properties thereof, and many have formulated hypotheses tests to evaluate changepoints. This study introduced another set of hypotheses, which involves sampling the length of the confidence intervals. With the estimated sampling distributions, the set of hypotheses proposed in (3.2) and (3.5) can now be tested, and naturally the power of the test can also be found. While this study has primarily investigated differences in mean, the same method can also be used to identify changepoints where the only slope is changed.

Oftentimes, using the asymptotic properties of the changepoint sampling distribution to form the theoretical distributions rely on a large number of data points. This was a motivation for the work by Hušková and Kirsch (2011). Their work used the bootstrap method to evaluate effects of differences in means on the power. In analyzing power by differences on the unit scale rather than the standard deviation scale, the results may not fully capture the effects of differences in mean and power. Thus, an analysis on the effect of differences in means in terms of standard deviation was necessary. 

This method can be applied to data streams, in which one can use the bootstrap to identify the most recent changepoint within a subset of data. Further extensions of this study, in order to be useful in data streams, can be done to evaluate the onlineness of the likelihood ratio method using the confidence interval length. The confidence interval length can also be further studied to see its efficacy in identifying new changepoints or outliers in data streams.

\end{document}